\documentclass[epj]{webofc}
\usepackage[varg]{txfonts}   
\usepackage{amsmath,amssymb}
\usepackage{graphicx}
\woctitle{MENU 2013}

\begin{document}

\title{Target and beam-target asymmetry measurements at MAMI}

\author{V.~L.~Kashevarov\inst{1,2}\fnsep\thanks{\email{kashev@kph.uni-mainz.de}} 
for Crystal Ball at MAMI, TAPS, and A2 Collaborations}

\institute{Institut f\"ur Kernphysik, Johannes Gutenberg-Universit\"at Mainz,
 D-55099 Mainz, Germany
\and Lebedev Physical Institute, 119991 Moscow, Russia}

\abstract{
Preliminary data for target and beam-target asymmetry for the reaction 
$\gamma p\to\pi^0\eta p$ obtained at the MAMI accelerator are presented.  
The data are compared to different theoretical predictions.}

\maketitle

\section{Introduction}
Main features of $\pi^0\eta$ photoproduction on protons were studied in works 
\cite{Doring,HornEPJ,Horn1940,FKLO,Anis12,W_LM}, 
were the reaction mechanism was explained through the dominance of the $D_{33}$ wave with
a small admixture of positive parity resonances and an insignificant Born terms 
contribution. 
Further steps toward investigation of this reaction are related to polarisation
observables that are especially sensitive to small reaction amplitudes. There are 64 
polarization observables for double pseudoscalar meson photoproduction. Different 
relationships among these observables and their symmetries decrease this value to 15
independent quantities \cite{RobertsOed}. Some of these were already measured and analyzed       
\cite{Ajaka,Gutz1,Gutz2,Aphi,DorMeiss}.

In the present work we measured at the first time the target and the beam-target 
asymmetries. 
For the totally exclusive five-fold cross section there are two independent transversal 
target asymmetries ($P_x$ and $P_y$) and two independent circular beam-target asymmetries 
($P_x^\odot$ and $P_y^\odot$) \cite{FiAr11}. 
The observables $P_y$ and $P_x^\odot$ integrated over the phase space of two from three final 
state particles are equivalent to $T$ and $F$ asymmetries of single pseudoscalar meson 
photoproduction. 
\section{Experimental setup and data analysis}
\begin{figure}
\begin{center}
\resizebox{0.48\textwidth}{!}{%
\includegraphics{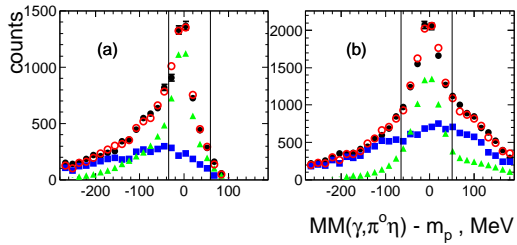}}
\caption{Carbon background subtraction. See text for an explanation.}
\label{fig1}
\end{center}
\end{figure}
The experiment was performed at the MAMI C accelerator in Mainz \,\cite{MAMIC} 
using the Glasgow-Mainz tagged photon facility \,\cite{TAGGER}. 
In the present measurement the longitudinally polarized electron beam with energy 
of 1557 MeV and polarization degree of 80\% was used. 
The longitudinal polarization of electrons is transferred to circular
polarization of the photons during the bremsstrahlung process in a radiator.
The reaction $\gamma p\to \pi^0\eta p$ was measured using the Crystal Ball \,\cite{CB} 
as the central spectrometer and TAPS \,\cite{TAPS} as a forward spectrometer.
The solid angle of the combined Crystal Ball and TAPS detection system is nearly 
$97\%$ of $4\pi$ sr. 

The experiment requires transversely polarized protons, which were provided by a
frozen-spin butanol ($C_4H_9OH$) target.
A specially designed $^3He/^4He$ dilution refrigerator was built 
in cooperation with JINR (Dubna).
For transverse polarization a 4-layer saddle coil was installed as the holding magnet,
which operated at a field of 0.45 Tesla.
The target container with 2-cm length and 2-cm diameter was filled by 2-mm diameter butanol
spheres with the filling factor of around $60\%$.
The average proton polarization during beam time period May-June 2010 and April 2011
was $70\%$ with relaxation times of around 1500 hours.
More details about the target are given in Ref.\,\cite{Thomas}.

The event-selection procedure was similar to the one that was described 
in Ref. \cite{PiEtaEPJA}, where meson pairs were identified via
their decay into 2 photons. For the case of 4 detected photons the best solution for the
meson pair was found using the $\chi^2$ minimization. This was then followed by
application of the invariant and missing mass cuts providing good identification of the
reaction.

Using the butanol target has an essential disadvantage because of an additional
background coming from the reactions on $^{12}C$ and $^{16}O$. Detection of the outgoing
protons and application of the coplanarity cut suppress the background significantly, but
the effect of this procedure is still insufficient. To subtract the residual
background events we used the results of the analysis of $\pi^0\eta$ photoproduction on
the carbon and the liquid hydrogen targets, which were fitted to the butanol data.
Because of the magnitude and the shape of the background depend on the initial beam 
energy and momenta of the final particles, the background subtraction procedure was 
performed for each bin, where the asymmetries were measured.
The procedure of the background subtraction is illustrated on Fig.\,\ref{fig1} for 
two different examples, which are typical for the presented data analysis. 
Missing mass spectra for the reaction $\gamma p\to \pi^0\eta p$ with the butanol target 
are shown on Fig.\,\ref{fig1} (a) and (b) by the full black circles. 
Spectra measured with the hydrogen and carbon targets are presented on the same plots 
by the green triangles and the blue squares correspondingly. Their absolute values were 
fitted to the butanol data. The red opened cycles, representing the sum of the hydrogen and 
carbon contribution, is our result of the fit. To minimize the uncertainties of the 
background subtraction, the procedure was used only within the missing mass energy interval, 
which is indicated on the plots by the vertical solid lines.

The systematic uncertainties come mainly from the determination of the proton
polarization degree (4\%), circular photon beam polarization degree (2\%), and the
background subtraction procedure (3-4\%) and were estimated to be less than 6\%. 
In order to reduce the systematic errors coming from target and detector conditions, 
the proton polarization direction was regularly reversed during experiment.
\section{Preliminary results}
\begin{figure*}
\begin{center}
\resizebox{0.9\textwidth}{!}{%
\includegraphics{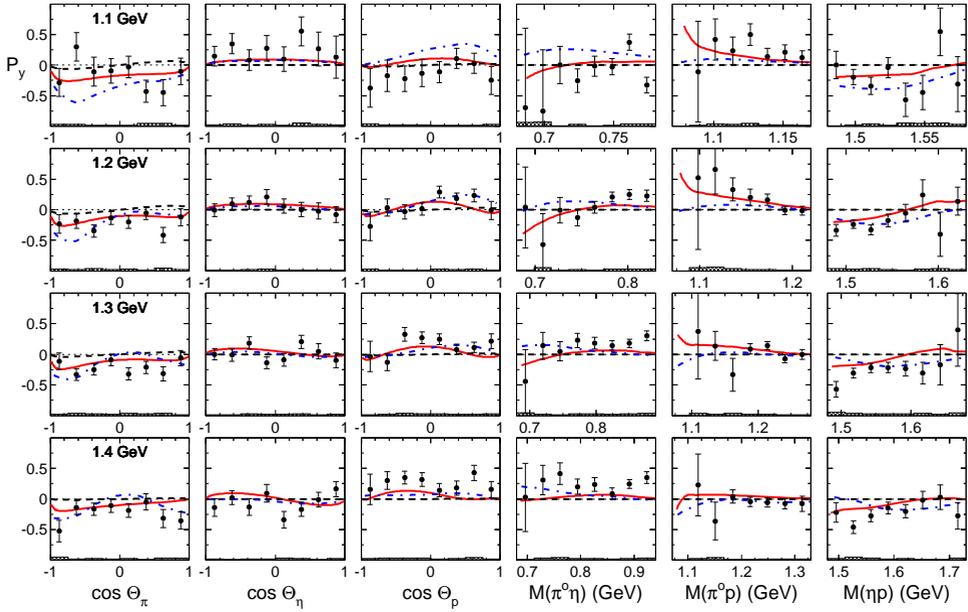}}
\caption{Angular and invariant mass distributions for the target asymmetry $P_y$.
Our preliminary data are shown by filled circles.
Solid curves show the prediction of the isobar model \protect\cite{FKLO}. Dashed
curves include only $3/2^-$ wave.
Predictions of the Bonn-Gatchina model \protect\cite{Anis12} are shown by 
dashed-dotted curves.
Histograms below are systematic errors.
The energy labels on the left panels indicate the central photon energy of 100 MeV bins.}
\label{fig2}
\end{center}
\end{figure*}
\begin{figure*}
\begin{center}
\resizebox{0.9\textwidth}{!}{%
\includegraphics{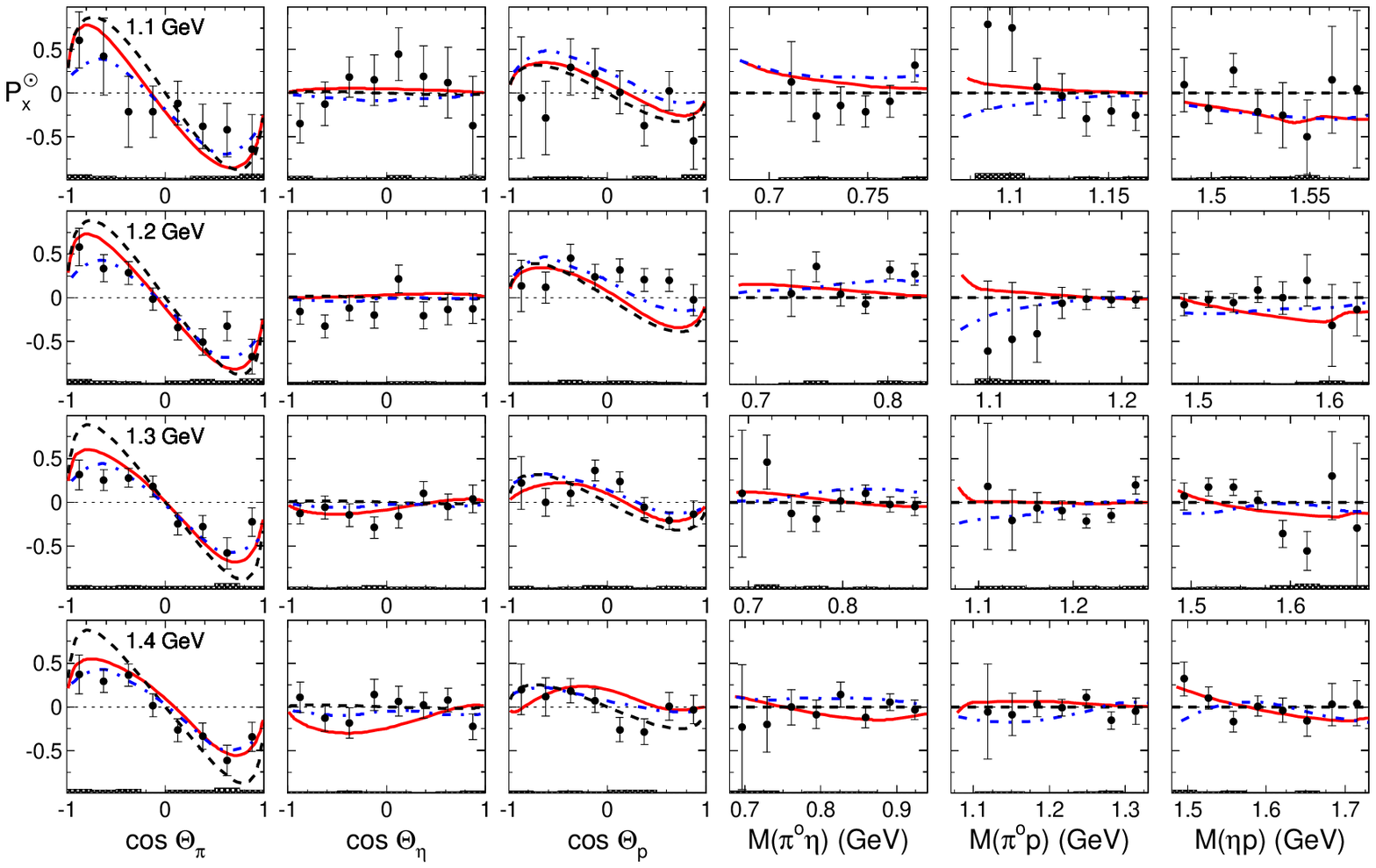}}
\caption{Same as in Fig.\,\protect\ref{fig2} for the beam-target asymmetry
$P_x^\odot$.}
\label{fig3}
\end{center}
\end{figure*}
Figs.\,\ref{fig2} and \ref{fig3} show our preliminary results together with
different theoretical predictions. 
The solid and dashed lines are predictions of the isobar model from Ref.\,\cite{FKLO}.
In this model the reaction amplitude contains the sum of $s$-channel Breit-Wigner
resonances with the total spin $J\leq 5/2$ and the Born terms. 
Parameters of the resonances were fitted to the experimental data \cite{PiEtaEPJA}. 
For the solid line we used the fit (I) which gives the best description of the 
measured linear beam asymmetry $\Sigma$ \cite{Ajaka,Gutz1}. 
Dashed line include only $\Delta 3/2^-$ resonances.

The dash-dotted line  shows predictions of the Bonn-Gatchina multichannel partial 
wave analysis \cite{Anis12} (solution BG2011-02). 
Within this approach the positions of resonances, their partial decay widths, and
relative strengths are fitted simultaneously to the data sets in different
channels, including single and double meson production as well as strangeness production. 
 
As we can see from Figs.\,\ref{fig2}-\ref{fig3}, both $P_y$ and $P_x^\odot$ demonstrate
more complicated behavior, than the one predicted by the single $\Delta 3/2^-$ model.
At the same time, the deviation is not large, thus indicating that the role of the states 
besides $\Delta 3/2^-$ remains restricted. 
Interference between $\Delta 3/2^-$ and the positive parity states $\Delta 1/2^+$ 
and $\Delta 3/2^+$ is responsible for the nontrivial angular and energy dependence of all 
asymmetries presented.

\section{Summary}
Preliminary data for the target and the beam-target asymmetry
of the cross section for $\gamma p\to\pi^0\eta p$ obtained with circular polarized
photons and transversally polarized protons were presented. The measurements were performed
at the MAMI C accelerator using the Crystal Ball/TAPS spectrometer. 
The polarization observables are sensitive to the contribution of the small components 
in the reaction amplitude. 
Obtained data could be usefully for further study of the partial wave content 
of $\pi\eta$ photoproduction.

This work was supported by the Deutsche Forschungsgemeinschaft (SFB 1044).

\end{document}